\documentclass[prl,aps,twocolumn,amsmath,amssymb,superscriptaddress,showpacs,showkeys]{revtex4-1}
\usepackage{graphicx}% Include figure files
\usepackage{dcolumn}% Align table columns on decimal point
\usepackage{bm}% bold math
\usepackage{float}
\usepackage[colorlinks=true, allcolors=blue, linktocpage=true]{hyperref}

\begin{document}
	
	\title{On-chip Hong-Ou-Mandel interference from separate quantum dot emitters in an integrated circuit}
		
	\author{\L{}ukasz Dusanowski}
	\email{lukaszd@princeton.edu}
	\affiliation{Technische Physik and W\"{u}rzburg-Dresden Cluster of Excellence ct.qmat, University of W\"{u}rzburg, Physikalisches Institut and Wilhelm-Conrad-R\"{o}ntgen-Research Center for Complex Material Systems, Am Hubland, D-97074 W\"{u}rzburg, Germany}
	\affiliation{currently at: Department of Electrical and Computer Engineering, Princeton University, Princeton, NJ 08544, USA}
		
	\author{Dominik K\"{o}ck}
	\affiliation{Technische Physik and W\"{u}rzburg-Dresden Cluster of Excellence ct.qmat, University of W\"{u}rzburg, Physikalisches Institut and Wilhelm-Conrad-R\"{o}ntgen-Research Center for Complex Material Systems, Am Hubland, D-97074 W\"{u}rzburg, Germany}
		
	\author{Christian Schneider}
	\affiliation{Technische Physik and W\"{u}rzburg-Dresden Cluster of Excellence ct.qmat, University of W\"{u}rzburg, Physikalisches Institut and Wilhelm-Conrad-R\"{o}ntgen-Research Center for Complex Material Systems, Am Hubland, D-97074 W\"{u}rzburg, Germany}
	\affiliation{Institute of Physics, University of Oldenburg, D-26129 Oldenburg, Germany}
		
	\author{Sven H\"{o}fling}
	\affiliation{Technische Physik and W\"{u}rzburg-Dresden Cluster of Excellence ct.qmat, University of W\"{u}rzburg, Physikalisches Institut and Wilhelm-Conrad-R\"{o}ntgen-Research Center for Complex Material Systems, Am Hubland, D-97074 W\"{u}rzburg, Germany}
	%\alsoaffiliation{SUPA, School of Physics and Astronomy, University of St Andrews, KY16 9SS St Andrews, UK}
		
	\date{\today}
	
%	\pacs{85.35.Be, 42.50.Ar, 42.82.-m, 42.50.Dv}
	
%\keywords{single photon source, integrated photonics, quantum dot, waveguides, two-photon interference}

	\begin{abstract}
		Scalable quantum photonic technologies require low-loss integration of many identical single-photon sources with photonic circuitry on a chip. Relatively complex quantum photonic circuits have already been demonstrated; however, sources used so far relied on parametric-down-conversion. Hence, the efficiency and scalability are intrinsically limited by the probabilistic nature of the sources. Quantum emitter-based single-photon sources are free of this limitation, but frequency matching of multiple emitters within a single circuit remains a challenge. In this work, we demonstrate a key component in this regard in the form of a fully monolithic GaAs circuit combing two frequency-matched quantum dot single-photon sources interconnected with a low-loss on-chip beamsplitter connected via single-mode ridge waveguides. This device enabled us to perform a two-photon interference experiment on-chip with visibility reaching 66\%, limited by the coherence of the emitters. Our device could be further scaled up, providing a clear path to increase the complexity of quantum circuits toward fully scalable integrated quantum technologies.
	\end{abstract}
	
	\maketitle
	
	Optical quantum computing and communication applications with single photons and linear optics rely critically on the quantum interference of two photons on a beamsplitter~\cite{kok2010introduction}. This process, known as Hong-Ou-Mandel (HOM) effect, occurs when two identical single photons enter a 50:50 beamsplitter, one in each input port. When the photons are indistinguishable, they will coalesce into a two-photon Fock state~\cite{Hong1987}, in which they exit the same but random output port. This process underlines the simplest non-trivial path-entangled NOON state generation and introduces an optical non-linearity which is the base for the implementation of more-complex photonic gates and protocols.  
	
	Consequently, scalable optical quantum information technologies will require integrating many identical indistinguishable single-photon sources with reliable photonic circuits consisting of beamsplitters. Utilizing well-developed integrated photonics technology is particularly appealing in this regard, as it dramatically reduces the footprint of quantum devices. Furthermore, it allows controlling photon states with high fidelity due to the intrinsic sub-wavelength stability of the path-lengths, low-losses, and near-perfect mode overlap at an integrated beamsplitter for high-fidelity quantum interference~\cite{Bonneau2016,Dietrich2016a,Wang2019}. 
	
	Advances in integrated photonic technology allowed already realizations of relatively complex quantum circuits demonstrating CNOT-gate operation~\cite{Crespi2011,Carolan2015}, boson sampling~\cite{Crespi2013,Carolan2015}, quantum walks~\cite{Peruzzo2010}, some simple quantum algorithms~\cite{Politi2009,Carolan2015} and chip-to-chip quantum teleportation~\cite{Llewellyn2019}. A combination of integrated photonic circuits with spontaneous four-wave mixing photon sources has also been achieved~\cite{Silverstone2013,Wang2018,Llewellyn2019}. However, due to the used sources' probabilistic nature, their efficiency and scalability are intrinsically limited. Quantum emitters-based single-photon sources are free of this limitation and recently have been shown to outperform spontaneous four-wave mixing and down-conversion photon sources in simultaneously reaching high levels of photons indistinguishability and brightness~\cite{Ding2016,Somaschi2016,Unsleber2016,Aharonovich2016,Senellart2017}. Moreover, remote interference between two quantum emitters was already demonstrated using trapped ions~\cite{Beugnon2006,Maunz2007}, quantum dots~\cite{Patel2010,Flagg2010,Konthasinghe2012,Gold2014,Kim2016a,Reindl2017a,Ellis2018,Weber2019,zhai_quantum_2022,you_quantum_2021}, organic molecules~\cite{Lettow2010,duquennoy_real-time_2022} or vacancy centers in diamond~\cite{Bernien2012,Sipahigil2012,Sipahigil2014,stolk_telecom-band_2022}. The vast majority of those experiments have been performed in free space as proof-of-principle demonstrations. Performing similar experiments on-chip, taking advantage of the photonic circuit consisting of fully integrated quantum emitters and beamsplitter, has not been performed yet, and it is still a missing component towards scaling-up aforementioned quantum technologies.
	
	In this work, we demonstrate a crucial component in this regard in the form of a fully monolithic GaAs circuit combing two frequency-matched quantum dot single-photon sources interconnected with on-chip beamsplitter via single-mode ridge waveguides. This device enabled performing two-photon interference experiments on-chip with visibility limited by the coherence of our emitters. 
	
	\begin{figure*}
		\includegraphics[width=6.5in]{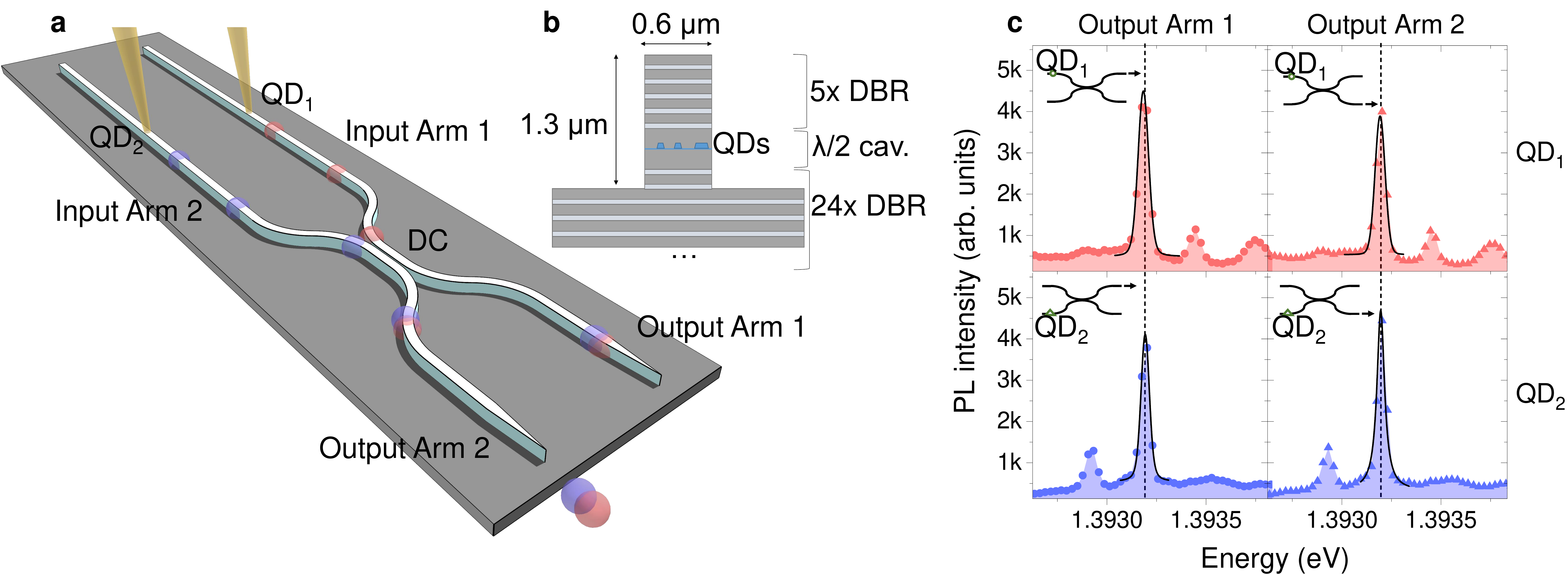}
		\caption{\label{fig:1}\textbf{On-chip two-photon interference circuit and beam splitting operation.} a,~Schematic representation of the photonic circuit based on a directional coupler (DC) interconnected with two input waveguides with coupled quantum dots and two output arms with inverted tapers for photons collection. b, Ridge waveguide cross-section with marked layer structure. c,~Demonstration of beam splitting operation for fabricated DC. The photoluminescence signal from QD$_1$ and QD$_2$ is recorded for Output Arms 1 and 2, respectively. QD$_1$ and QD$_2$ are frequency matched with a precision of 5~$\mu$eV.}
	\end{figure*}
	
	Our semiconductor photonic device is schematically presented in Fig.~\ref{fig:1}a and b. It is based on InAs/GaAs distributed Bragg-reflector ridge waveguides, which have been proven to facilitate high optical quality quantum dot single-photon sources~\cite{Dusanowski2019}. The central part of the device consists of the single-mode directional coupler (DC), which is the integrated optical analog of the bulk beamsplitter. In the two input arms of the DC, two frequency-matched quantum dots (QDs) are located. For single-photon generation, the QDs are excited non-resonantly from the top using two separated picosecond pulsed laser beams. Photons interfered on the DC are finally collected off the chip using inverse taper out-couplers. Spectral filtering and detection are performed off-chip using a monochromator and two superconducting single-photon detectors.
	
	To find two quantum dots with matching optical transition energies, the position of the excitation beam spot on each input arm of the DC was scanned using an automatized translation stage. Within such a scanning routine, we localized two matching emission lines at 1.3931~eV energy originating from the QDs located in two individual input arms of the DC and separated spatially by around 200~$\mu$m. In Fig.~\ref{fig:1}b, photoluminescence (PL) spectra from QD$_1$ and QD$_2$ recorded from DC output arms 1 and 2 are presented at a temperature of 4.5~K. Single well-resolved emission lines matching within 5~$\mu$eV fit precision are visible. Comparing amplitudes of QD$_1$ and QD$_2$ emission peaks visible within both output arms, a beam splitting ratio of 48:52 is derived (including uneven transmission through out-coupling arms - more details in Supplementary Section~8).
	
	To show that optical excitation of our QDs leads to the generation of single photons, we analyzed the photon emission statistics of separate QDs by performing second order-correlation experiments in Hanbury Brown and Twiss (HBT) configuration. For that purpose, QDs have been excited non-resonantly from the top by an 813~nm wavelength train of picosecond pulses at a repetition rate of 76~HMz. Photons emitted by the QDs were then coupled into the circuit input arm waveguides and guided into the directional coupler, where the signal was divided between two output arms. Next, photons were collected off-chip from the side of the sample using out-couplers and subsequently filtered spectrally by a monochromator (70~$\mu$eV width) and coupled into two single-mode fibres connected with superconducting single-photon detectors (SSPD). Finally, the photon correlation statistics were acquired by a multichannel picosecond event timer. Data have been recorded under excitation powers corresponding to half of the QD saturation intensity.
	
	\begin{figure}
		\includegraphics[width=3.3in]{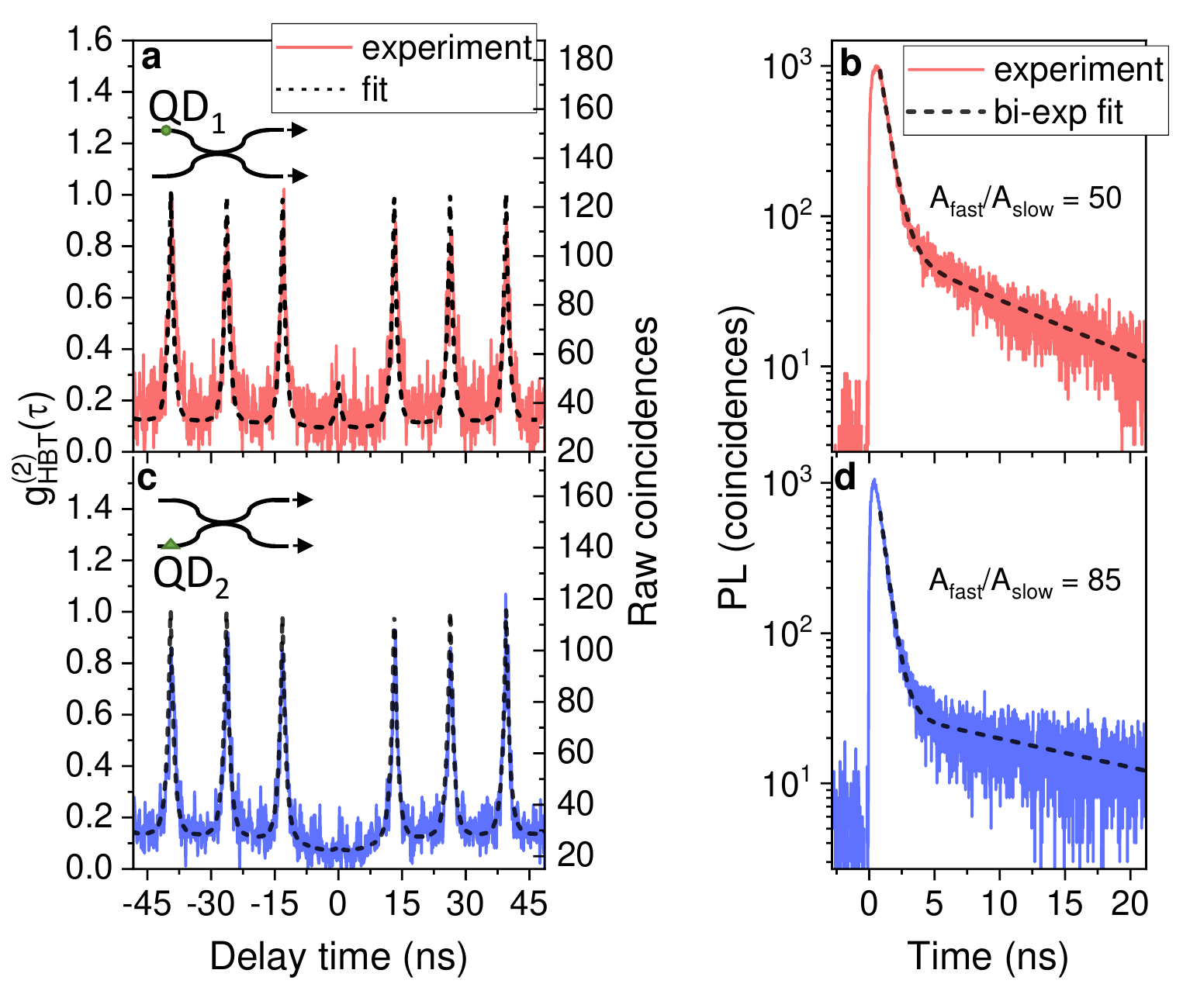}
		\caption{\label{fig:2}\textbf{Single-photon generation and emission dynamics.} a,c~Second order auto-correlation histograms of QD$_1$ and QD$_2$ emission under pulsed 76~MHz repetition rate excitation. Data have been recorded in HBT configuration using an on-chip beamsplitter. b,d~Time-resolved PL traces revealing bi-exponential decays with fast (slow) time constant of 720~ps (12~ns) and 600~ps (22~ns) for QD$_1$ and QD$_2$, respectively.}
	\end{figure}
	
	Fig.~\ref{fig:2}a and c present the second-order autocorrelation function $g^{(2)}_{HBT}(\tau)$ measurement recorded for each QD individually. In the case of both QDs, a clear suppression of the central peak counts is visible, proving single-photon emission. To quantitatively evaluate the probability of multi-photon emission, $g^{(2)}_{HBT}(0)$ values were calculated by integrating residual counts of the zero delay peak with respect to the neighboring six peaks, resulting in $g^{(2)}_{HBT}(0)=0.35\pm0.08$ and $g^{(2)}_{HBT}(0)=0.15\pm0.02$ for QD$_1$ and QD$_2$, respectively. In Fig.~\ref{fig:2}b and d, time-resolved photoluminescence traces of the QD$_1$ and QD$_2$ emission are shown. In this case, the repetition rate of the laser was reduced to 19~MHz using a pulse picker. Clear bi-exponential signal decays are visible, with a fast and slow time constant of 720$\pm$5~ps and 12$\pm$1~ns for QD$_1$ and 600$\pm$5~ps and 22$\pm$1~ns for QD$_2$. We attribute the fast decay to the spontaneous recombination of electron-hole pairs in QD ($T_1$) and the slow one, which corresponds to about 2\% (1.2\%) of the total QD$_1$ (QD$_2$) line intensity, is tentatively interpreted as the recapturing of the carriers by the QD. Using fit parameters obtained from the time-resolved experiments, $g^{(2)}_{HBT}(\tau)$ correlation histograms have been fitted with double-sided bi-exponential decay convoluted with 80~ps width Gaussian instrumental response function (black dashed lines). 
	
	\begin{figure*}
		\includegraphics[width=6.5in]{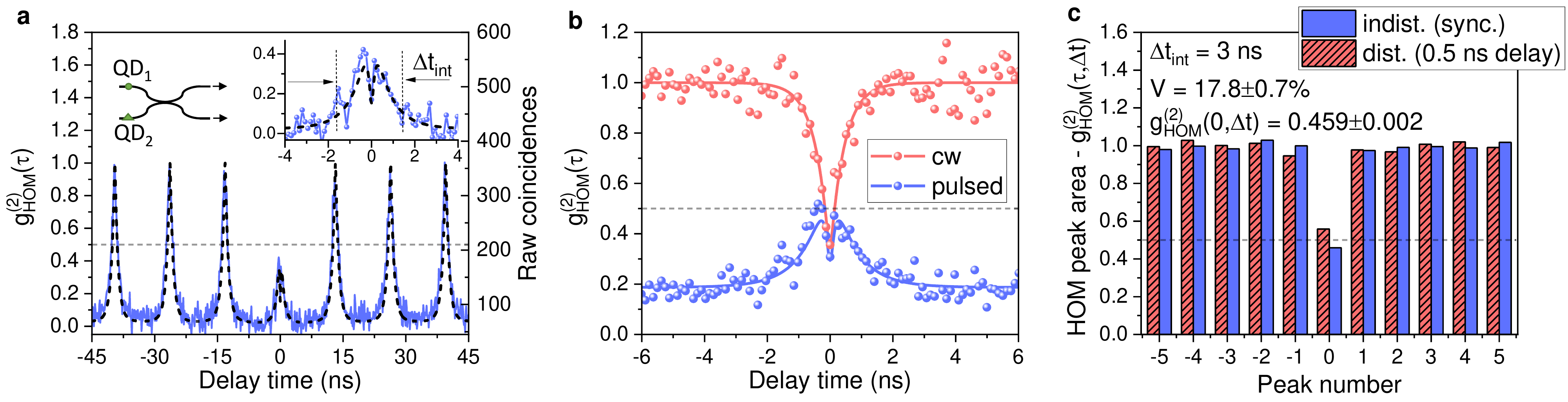}
		\caption{\label{fig:3}\textbf{ On-chip two-photon interference from separate quantum emitters.} a,~Two-photon Hong-Ou-Mandel interference measurement between QD$_1$ and QD$_2$ showing the normalized HOM coincidences versus the delay time. The central peak area is suppressed with respect to neighboring peaks. Inset: Magnified view of the central peak area. b,~Raw HOM interference measurement recorded under cw (red points) and pulsed (blue points) excitation. c,~Integrated counts of the central eleven peaks ($\Delta t$~=~3~ns integration window) of the HOM correlation in case of synchronized (blue bars) and 0.5~ns delayed (red bars) photons from QD$_1$ and QD$_2$. All presented data are recorded using an on-chip beamsplitter.}
	\end{figure*}
	
	%make it sounds better
	To demonstrate on-chip two-photon interference, the single QDs in both input arms of the DC are excited using the picosecond pulses. For that, the laser beam is divided into two independently controllable optical excitation axis and synchronized in advance to ensure optimal temporal overlap of emitted photons on the DC. It is performed by sending the emission from each QD separately through the on-chip DC and using time-resolved detection to eliminate the time delay difference between independently generated single photons. The same technique is used to introduce an intentional 0.5~ns time delay for reference measurements. The excitation laser powers for each QD are adjusted such that their emission intensities are the same (around half of QD$_1$ saturated intensity). As we utilize on-chip beam splitting operation using DC with single-mode inputs and outputs, we expect a very high spatial mode overlap of our interferometer. To test this, we send the continuous wave laser simultaneously into both DC input arms and record classical interference fringes with 98$\pm$1\% visibility. In earlier experiments performed on ridge waveguide structures, we observed that the QD emission couples into the well-defined transverse-electric mode of the WG with close to unity degree of linear polarization~\cite{Dusanowski2019}. In the case of the investigated device, the polarization of the emitted photons was analyzed after passing the whole circuit consisting of bend regions, DC itself, and out-couplers (more details in Supplementary Section~6). We found that for both QDs, the degree of polarization is above 95\%, suggesting optimal polarization alignment for interference experiments.  
	
	Within the above-mentioned prerequisites, the two-photon interference should be mainly limited by the coherence of our single-photon emitters. To get access to the coherence times of our QDs, we performed a high-resolution measurement of the emission linewidths using a scanning Fabry-Perot interferometer. We extract the full-width at half-maximum of 13.5$\pm$2.5~$\mu$eV and 3.0$\pm$0.2~$\mu$eV by Lorentzian fit for QD$_1$ and QD$_2$, respectively (see Supplementary Section~7). The coherence times calculated based on observed broadenings are $T_2^{QD_1}$=~100$\pm$20~ps and $T_2^{QD_2}$=~440$\pm$30~ps. As the measurements are performed on the tens of seconds timescale, we speculate that recorded coherence times might be limited by charge and spin noise~\cite{Kuhlmann2013,Makhonin2014}. Following Ref.~\cite{Kambs2018}, we calculated the expected interference visibility of our two independent emitters and derived the theoretical visibility in the range of $V_{theory}=$~10-15\%.  
	
	Figure~\ref{fig:3}a shows two-photon interference data in the form of second-order HOM cross-correlation between photons exiting the two output arms of the on-chip beamsplitter. The height of the central peak is clearly below the half intensity of the neighboring peaks, proving that photons emitted by two separate QDs indeed interfere on the DC. Another interference signature is the presence of the coincidences dip superimposed on the central peak around zero time delay. The depth of this dip constitutes to the interference events where photons arrive simultaneously at the DC, giving rise to the narrow-time window post-selected coalescence. In our case, the exact value of $g^{(2)}_{HOM}(\tau)$ at $\tau=$~0 is equal to 0.17 in the case of background-corrected data and 0.31 for as-measured data. The same type of time post-selected interference can be observed for cw HOM correlations. Figure~\ref{fig:3}b shows the non-corrected cw, and pulsed HOM interference histograms overlapped on each other (corresponding cw $g^{(2)}_{HBT}(\tau)$ graphs are shown in Supplementary Section~10). Similar to the pulsed case, the cw correlation shows clear suppression of coincident counts at zero time delay, with time post-selected $g^{(2)}_{HOM}(0)$ of 0.35, close to the pulsed as-measured value of 0.31. 
	
	To evaluate photons full wave-packet interference probability (non-post-selected), we calculate the pulsed HOM correlation central peak area normalized by the average area of the neighboring six peaks. For integration window $\Delta t$ of 3~ns, we obtain $g^{(2)}_{HOM}(0,\Delta t)=$~0.459$\pm$0.002 for background corrected data and $g^{(2)}_{HOM}(0,\Delta t)=$~0.587$\pm$0.002 for raw data, where uncertainty is based on the standard deviation of non-central peaks areas. In the case of background-corrected data, we reach a value below the 0.5 classical limit. It needs to be noted that derived $g^{(2)}_{HOM}(0,\Delta t)$ and $g^{(2)}_{HOM}(0)$ values are partially influenced by the non-zero multi-photon emission extend observed in HBT measurements.
	
	As it has been recently pointed out in Ref.~\cite{Jons2017,Weber2018}, to estimate two-photon interference visibility for remote emitters correctly, it is necessary to perform reference HOM measurements for distinguishable photons, due to the possible blinking effect. Since within the fabricated circuit polarization rotation is impossible to unambiguously confirm the two-photon interference and properly evaluate visibility, photons were made distinguishable by introducing a 0.5~ns time delay between excitation pulses. Such delay should be sufficient to lose the temporal photons overlap on the DC within the emitters coherence times and record reference data.
	
	Figure~\ref{fig:3}c demonstrates the normalized histogram of the central eleven peaks areas ($\Delta t$~=~3~ns)~of the HOM second-order cross-correlation in case of synchronized - indistinguishable (red bars) and 0.5~ns delayed - distinguishable (grey bars) photons. The central peak area in the case of unsynchronized photons is equal to $g^{(2)}_{HOM_{d}}(0,\Delta t)=$~0.558$\pm$0.002, which is slightly above the theoretically expected 0.5 value. We relate this discrepancy with non-zero multi-photon emission extend. Finally, we calculate remote sources two-photon interference visibility $V$ following $V=[g^{(2)}_{HOM_{d}}(0,\Delta t)-g^{(2)}_{HOM}(0,\Delta t)]/g^{(2)}_{HOM_{d}}(0,\Delta t)$, resulting in $V=$~17.8$\pm$0.7\% for background corrected data. This value is relatively close to the theoretically expected visibility and even partially exceeds it, suggesting that it is limited solely by the coherence of the emitters (more details Supplementary Section~9). Using background corrected data for the pulsed case, give post-selected visibility of $V'_{p}=$~66\%. The probability of the time post-selected interference is known to depend on the ratio of the emitters coherence times to the setup timing resolution~\cite{Kiraz2004,Patel2010,Flagg2010}, thus possibly even higher $V'$ values could be potentially achieved with faster detectors.
	
	\begin{figure*}
	\includegraphics[width=6.5in]{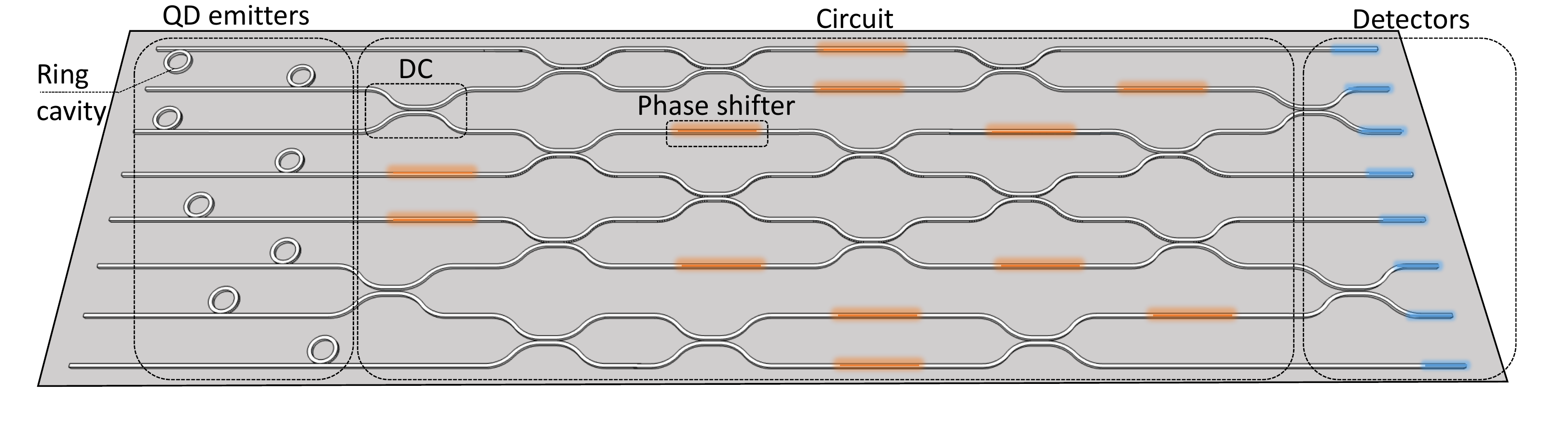}
	\caption{\label{fig:4}\textbf{Envisioned fully integrated quantum photonic circuit.} Draft of the possible circuit design with multiple quantum dot-based single photon sources coupled to ring cavities, interconnected with ridge waveguides, directional couplers, phase shifters, and superconducting detectors.}
	\end{figure*}
	
	While our results provide clear scientific evidence for on-chip generation and interference of on-demand single photons in a circuit, the recorded visibility values need to be improved for future practical applications. In the current device architecture, the indistinguishability of interfered photons is limited by $T_2/(2T_1)$ of particular QDs. We propose a few strategies for improving this ratio. Firstly, the QD charge environment could be stabilized via passivation~\cite{Press2010}, weak optical illumination~\cite{Majumdar2011}, or gating~\cite{Somaschi2016, Liu2018}. At the same time, by embedding QDs into optical cavities, the Purcell effect might be used to enhance the radiative emission rate $1/T_1$~\cite{Ding2016,Somaschi2016,Unsleber2016,Liu2018,Iles-Smith2017}. Recently, we demonstrated a QD circuit with ring cavities allowing to significantly increase the QD coupling efficiency into the WG mode and decrease the $T_1$ below 200~ps~\cite{dusanowski_purcell-enhanced_2020}. Finally, by applying a resonant excitation, the photons emission time-jitter could be minimized, and strong suppression of multi-photon emission events achieved~\cite{Ding2016,Somaschi2016,Unsleber2016,Liu2018,Dusanowski2019,dusanowski_purcell-enhanced_2020}. Within such circuit and excitation improvements, two-photon interference with near-unity visibility seems to be within reach. 
	
	To realize circuits combining multiple QD sources coupled to cavities, deterministic fabrication technologies such as in-situ electron-beam lithography or imaging will be required. This will allow to preselect emitters with identical spectral characteristics, build cavities around them and combine them within a single functional photonic circuit. In principle, since QD imaging could be performed in an automatized manner, a very large number of emitters could be combined on a single chip. At such a stage of complexity, separate control over QD emission energies might also be desired. This could be directly implemented by a local laser drive via AC Stark effect~\cite{dusanowski_all-optical_2022} or adapting the circuit for the electric field~\cite{Patel2010,Liu2018}  and strain~\cite{Flagg2010,Beetz2013,Elshaari2018,Moczala-Dusanowska2019} control. Ultimately, a practical quantum photonic chip will require the presence of additional functionalities such as single-photon detectors and phase-shifter. Fortunately, the GaAs circuits are compatible with superconducting detectors technology~\cite{Sprengers2011,Reithmaier2015,Schwartz2018} and thanks to the large $\chi^2$  nonlinear coefficient of the GaAs, electro-optical phase shifters have already been demonstrated~\cite{Wang2014a}. Such an envisioned fully functional QDs-GaAs circuit is shown schematically in Fig.~\ref{fig:4}.
	
	In conclusion, we have shown that two identical QD single-photon sources can be integrated monolithically in a waveguide circuit and made to interfere with visibility limited by the coherence of those sources. We pointed out the potential strategies to improve the QDs performance by employing deterministic fabrication and cavity enhancement. The implemented integrated system could be potentially further extended to facilitate more complex circuits and fully on-chip operation. Results shown in this article, along with a clearly outlined path for future improvements, take us one step closer to scalable integrated quantum circuits based on quantum emitters capable of generating and manipulating large photonic states.
	\\ \textbf{Methods}
	\\ \textbf{Sample description.}
	To fabricate our integrated single-photon source waveguide device, we use a semiconductor sample that contains self-assembled In(Ga)As QDs grown by the Stranski-Krastanow method at the center of a planar GaAs microcavity. The lower and upper cavity mirrors contain 24 and 5 pairs of Al$_{0.9}$Ga$_{0.1}$As/GaAs $\lambda$/4-layers, respectively, yielding a quality factor of $\sim$200. A $\delta$-doping layer of Si donors with a surface density of roughly $\sim$10$^{10}$~cm$^{-2}$ was grown 10~nm below the layer of QDs to dope them probabilistically. To fabricate ridge waveguides devices, the top mirror layer along with the GaAs cavity is etched down, forming the ridge with a width of $\sim$0.6~$\mu$m and a height of $\sim$1.3~$\mu$m. The cross-section of the WG with layer structure is shown in Figure~\ref{fig:1}b (see also Supplementary Section~1). Ridges have been defined by e-beam lithography and reactive ion etching. After processing, the sample was cleaved perpendicularly to the WGs, around 30~$\mu$m away from the tapered out-coupler edges to get clear side access.  
	\\ \textbf{Integrated circuit design}
	We designed and fabricated GaAs directional couplers with different coupling lengths and gaps. A directional coupler with a near 50:50 coupling ratio at around 1.3931~eV was obtained when the gap distance was set to 120~nm and the coupling length to 30~$\mu$m (see Supplementary Section~2 for layout scheme). The total length of the device was about 1 mm, including four S-bends with a radius of 60~$\mu$m and the input/output waveguides.
	\\ \textbf{Experimental setup.}
	For all experiments, the sample is kept in a low-vibrations closed-cycle cryostat (attoDry800) at temperatures of $\sim$4.5~K. The cryostat is equipped with two optical windows allowing for access from the side and the top of the sample. A spectroscopic setup consisting of two independent perpendicularly aligned optical paths is employed (see Supplementary Section~3 for more details). QDs embedded into WGs are excited from the top through a first microscope objective with NA~=~0.26, while the emission signal is detected from a side facet of the WG with a second objective with NA~=~0.4. The photoluminescence signal, simultaneously collected from both output arms of the DC, is then passed through a Dove prim to rotate the sample image plane from a horizontal into a vertical direction to fit the monochromator slit orientation. For PL analysis, the signal is then spectrally dispersed by a 75~cm focal length monochromator and focused on a low-noise liquid-nitrogen-cooled CCD camera (around 40~$\mu$eV spectral resolution), allowing to resolve signal from both DC output arms spatially. For HBT and HOM experiments, the monochromator serves as a spectral filter with 70~$\mu$eV width, and the signal from both DC outputs is introduced into separate single-mode optical fibres connected with superconducting single-photon counting detectors (30~ps time-response).
	\\ \textbf{Integrated beamsplitter visibility.} 
	To test the classical visibility of the DC device, we simultaneously send the continuous wave laser light tuned to the energy of QDs transitions, using circular reflectors placed on the ends of the input arms waveguides (see Supplementary Section~8). The power of the laser coupled into both arms was adjusted such that the intensity from both input arms was the same. Next, we focused on the signal passing through DC and out coupled from one output arm. We observed intensity modulation in the function of time, related to small path-length difference fluctuation, allowing us to see interference pattern and calculate the interferometer visibility. In the case of the investigated device, the classical visibility of 98$\pm$1\% was extracted.
	\\ \textbf{Correlation histograms analysis.}
	For the time post-selected visibility $V'$ analysis, we assume that for distinguishable photons $g^{(2)}_{HOM_{d}}(0)$ is equal to 0.5, as reference measurement is not possible. It allows to calculate $V'$ according to $V'=[0.5-g^{(2)}_{HOM}(0)]/0.5$. Data from Fig.3b lead to raw visibility of $V'_{cw}=$~30\% and $V'_{p}=$~38\% for cw and pulsed excitation, respectively. For $g^{(2)}_{HBT}(\tau)$ and $g^{(2)}_{HOM}(\tau)$ correlation functions evaluation we take into account a presence of the time-independent background offset in recorded histograms (it constitutes to around 15-20\% of the coincidences), which we relate to the dark counts of the SSPDs (100-500~cps). Non-background-corrected HOM graphs can be found in Fig.~\ref{fig:3}c and the Supplementary Section~9.      
	\begin{acknowledgments}
		The authors thank Silke Kuhn for fabricating the structures. \L{}.D. acknowledges the financial support from the Alexander von Humboldt Foundation. We acknowledge financial support by the German Ministry of Education and Research (BMBF) within the project "Q.Link.X" (FKZ: 16KIS0871). We are furthermore grateful for the support by the State of Bavaria.
	\end{acknowledgments}

%	\begin{suppinfo}
%	The Supporting Information is available free of charge:\\
%	Sample structure; Integrated circuits layout; Optical Set-Up; Power-resolved PL; Waveguide transmission losses; Polarization resolved PL; Emitters transition linewidths; Directional coupler characteristics; Raw HOM correlation data; Single photon emission under cw excitation. 	
%	\end{suppinfo}

%	\textbf{Author Contributions}
%	C.S. and S.H. designed and grew the wafer. \L{}.D. designed the structure and guided the fabrication process. \L{}.D. established an experimental setup, carried out the optical experiments, analyzed and interpreted the experimental data. D.K. and \L{}.D. performed the numerical simulations. S.H. and C.S. guided the work. \L{}.D. wrote the manuscript with input from all authors.
	
%	\textbf{Data availability}
%	The data that support the findings of this study are available from the corresponding author on reasonable request.
	
	\bibliography{bib-manuscript}% Produces the bibliography via BibTeX.content...

\end{document}

\title{Supporting Information:\\ On-chip Hong-Ou-Mandel interference from separate quantum dot emitters in an integrated circuit}
	
	\author{\L{}ukasz Dusanowski}
	\email{lukasz.dusanowski@uni-wuerzburg.de}
	\affiliation{Technische Physik and W\"{u}rzburg-Dresden Cluster of Excellence ct.qmat, University of W\"{u}rzburg, Physikalisches Institut and Wilhelm-Conrad-R\"{o}ntgen-Research Center for Complex Material Systems, Am Hubland, D-97074 W\"{u}rzburg, Germany}
	\affiliation{currently at: Department of Electrical and Computer Engineering, Princeton University, Princeton, NJ 08544, USA}
	
	\author{Dominik K\"{o}ck}
	\affiliation{Technische Physik and W\"{u}rzburg-Dresden Cluster of Excellence ct.qmat, University of W\"{u}rzburg, Physikalisches Institut and Wilhelm-Conrad-R\"{o}ntgen-Research Center for Complex Material Systems, Am Hubland, D-97074 W\"{u}rzburg, Germany}
	
	\author{Christian Schneider}
	\affiliation{Technische Physik and W\"{u}rzburg-Dresden Cluster of Excellence ct.qmat, University of W\"{u}rzburg, Physikalisches Institut and Wilhelm-Conrad-R\"{o}ntgen-Research Center for Complex Material Systems, Am Hubland, D-97074 W\"{u}rzburg, Germany}
	\affiliation{Institute of Physics, University of Oldenburg, D-26129 Oldenburg, Germany}
	
	\author{Sven H\"{o}fling}
	\affiliation{Technische Physik and W\"{u}rzburg-Dresden Cluster of Excellence ct.qmat, University of W\"{u}rzburg, Physikalisches Institut and Wilhelm-Conrad-R\"{o}ntgen-Research Center for Complex Material Systems, Am Hubland, D-97074 W\"{u}rzburg, Germany}
	%\affiliation{SUPA, School of Physics and Astronomy, University of St Andrews, KY16 9SS St Andrews, UK}
	
	\date{\today}% It is always \today, today,
	%  but any date may be explicitly specified

	\maketitle
	
	\renewcommand\thefigure{S\arabic{figure}}   
	
	%sample details
	\section{S1: Sample structure} 
\begin{figure}[h]
		\includegraphics[width=4.5in]{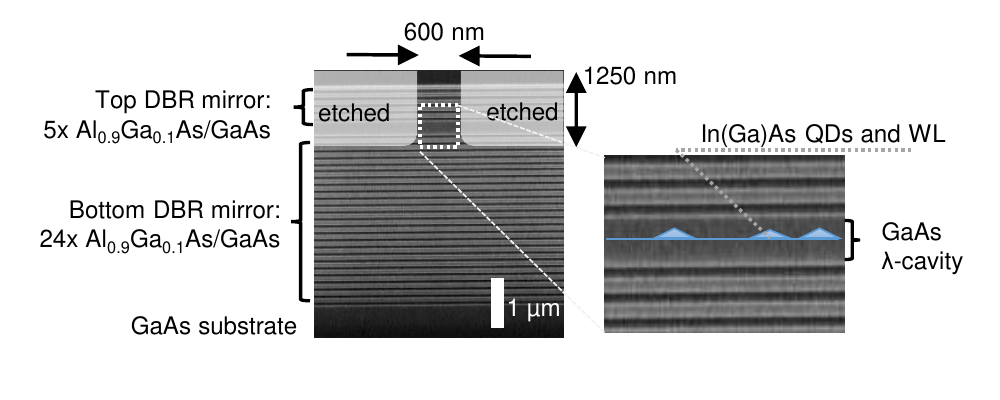}
		\caption{\label{fig:sample}Planar sample scanning electron microscope cross-section image with visible layers and schematically marked areas for etching. The quantum dot layer is placed inside a center of a $\lambda$ cavity sandwiched between two distributed Bragg Reflectors consisting of the 5/24 alternating $\lambda$/4-thick layers of Al$_{0.9}$Ga$_{0.1}$As and GaAs.  
		}
	\end{figure}

	The full layer structure of the sample is shown in Figure~\ref{fig:sample}.
	
	\section{S2: Integrated circuit layout} 
\begin{figure}[h]
		\includegraphics[width=6.5in]{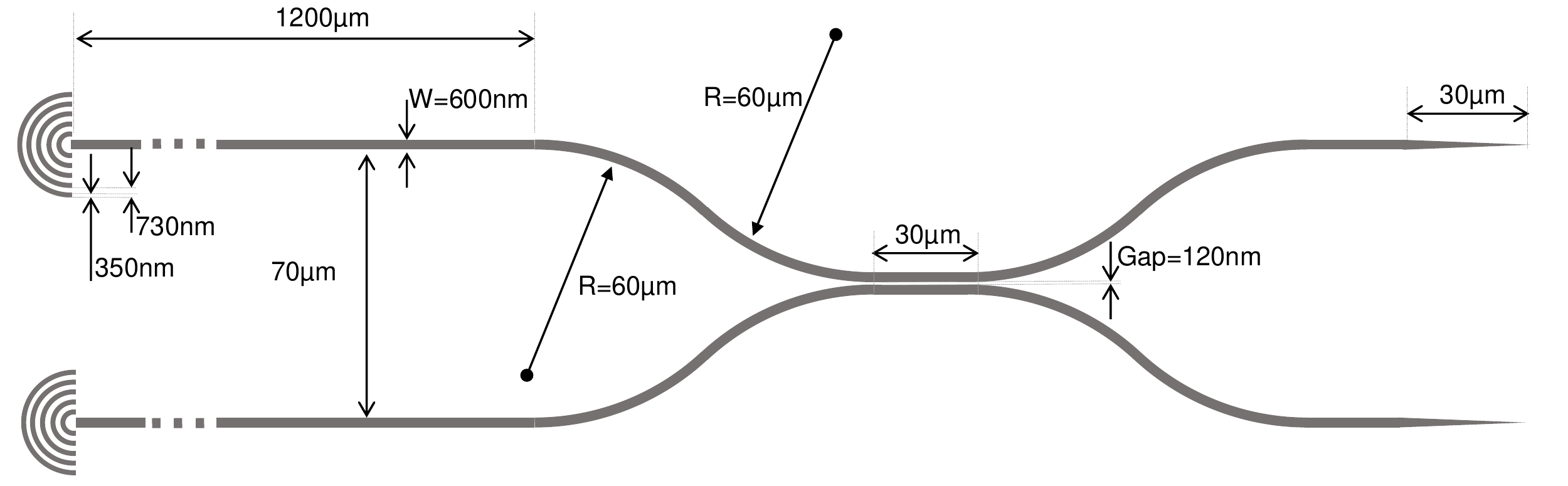}
		\caption{\label{fig:circuit}Integrated circuit layout. Scheme of the fabricated GaAs directional coupler including circular Bragg reflectors located at the ends of the input WG arms and inverse taper out-couplers at the end of the output arms. Bending regions are based on circular profiles with a radius of 60~$\mu$m. 
		}
	\end{figure}

\begin{figure}[h]
		\includegraphics[width=5.5in]{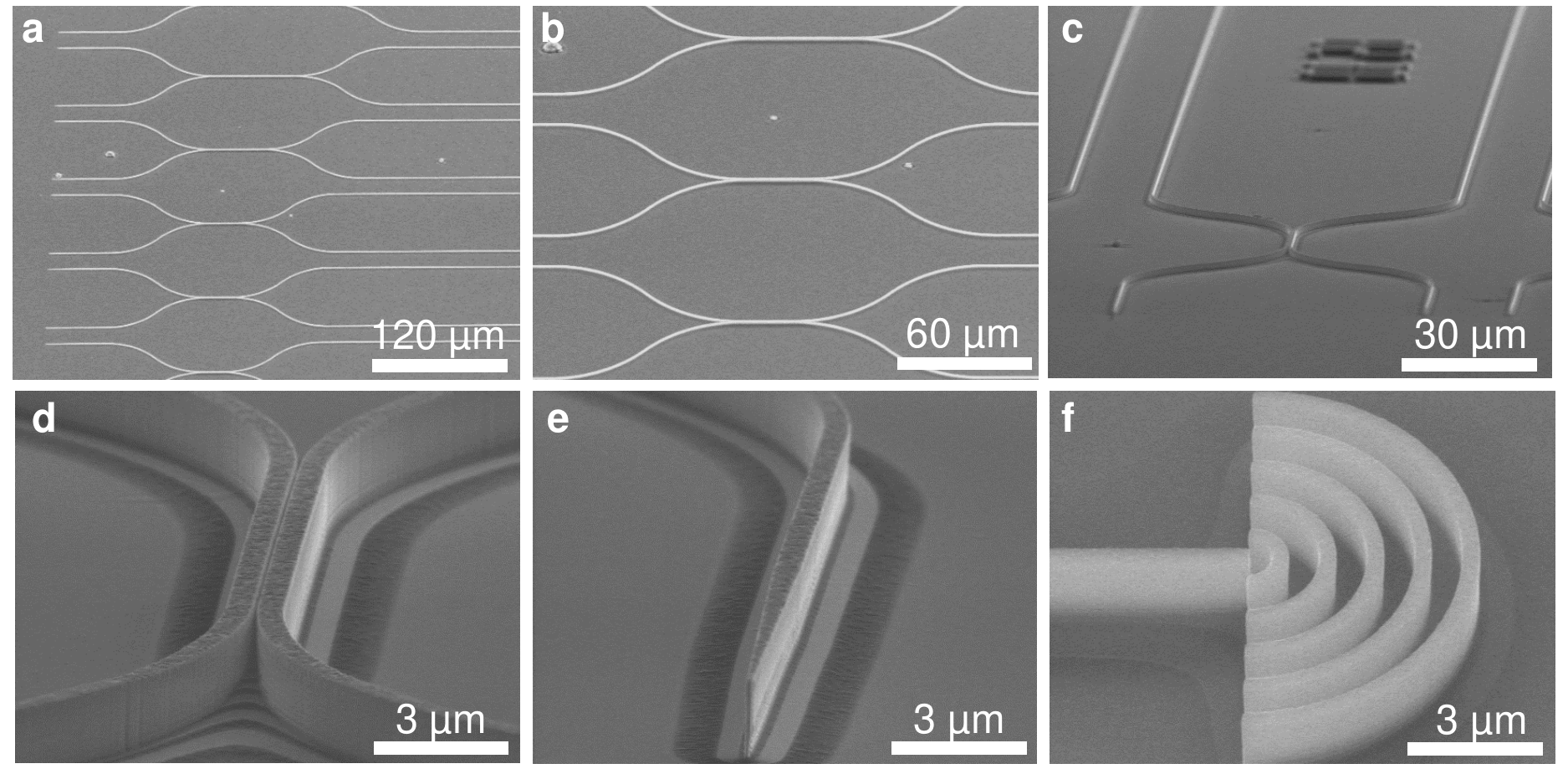}
		\caption{\label{fig:SEM}Scanning electron microscope images of the fabricated devices. a,b,~The DCs with different coupling lengths. c,d,~The DC with 30~$\mu$m length coupling region. e,~An inverse taper outcoupler. f,~A circular Bragg grating reflector.
		}
	\end{figure}

Figure~\ref{fig:circuit} shows the layout scheme of the fabricated GaAs device. It is based on 600~nm width single-mode ridge waveguides. The central part of the circuit is a directional coupler (DC) formed by two WGs separated by a 120~nm gap along a 30~$\mu$m long coupling region. WGs were brought together using two circular bend regions with a radius of 60~$\mu$m. Waveguides on the right end of the circuit are terminated by inverse taper (30~$\mu$m length) out-couplers, minimizing reflection and optimized for better light extraction out of the chip. The left side of the DC consists of 1.2~mm long straight WG sections designated for searching two quantum dots with the same transition frequencies. WGs on the left side of the circuit are terminated with circular Bragg grating mirrors optimized for increased reflectivity (around 80\% expected at 900~nm). Figure~\ref{fig:SEM} shows the scanning electron microscope images of the fabricated integrated circuits.   
	
	\begin{figure}[h]
		\includegraphics[width=6.5in]{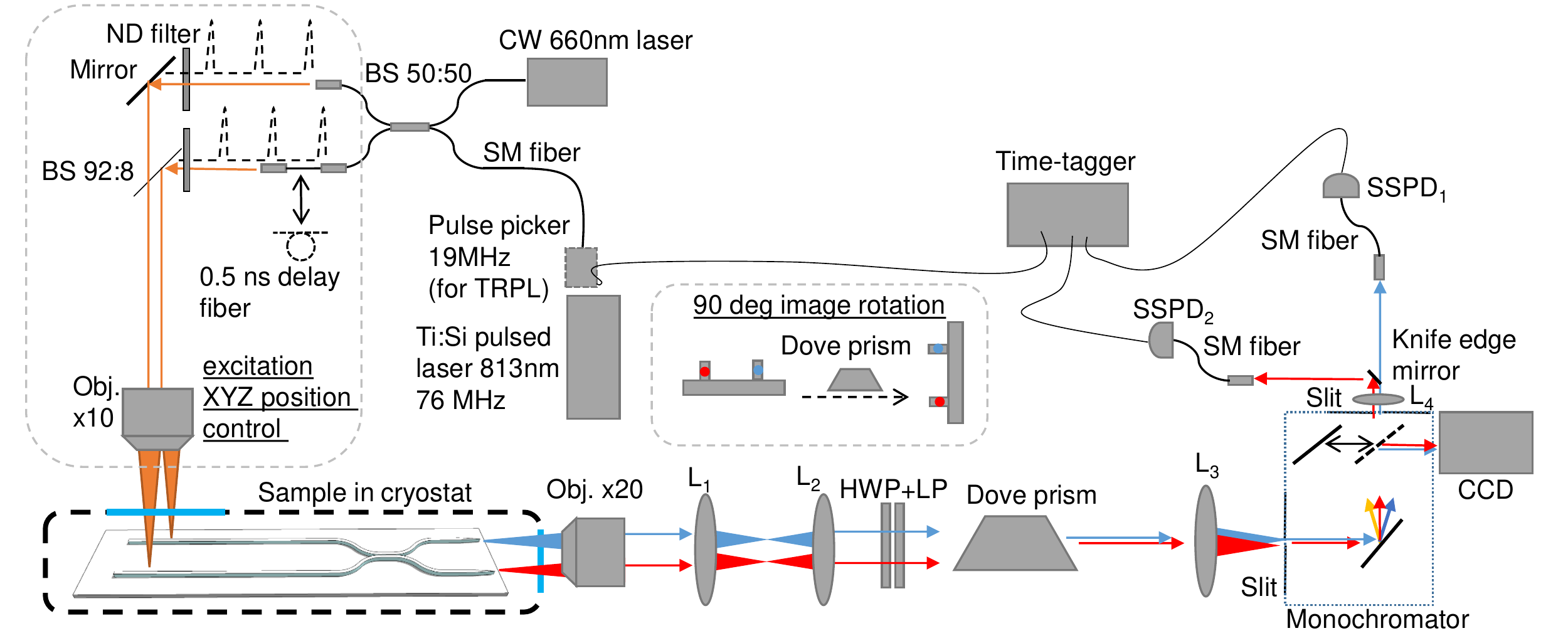}
		\caption{\label{fig:setup}Optical setup. Scheme of the experimental configuration used for top excitation (blue path) and side detection (red path) photoluminescence and resonance fluorescence measurements. In the case of two-photon interference experiments, a QD was excited twice every laser pulse cycle with a delay of 3~ns, and the subsequently emitted photons, spatially and temporally overlapped in an unbalanced Mach-Zehnder interferometer (dashed lines) utilizing polarization maintaining (PM) fibers and beam-splitters (BS). For signal detection, two avalanche photo-diodes (APD) with 350~ps response time were used. For polarization control in free space, a half-wave-plate (HWP) combined with a linear polarizer (LP) was used, while for polarization rotation (PR) in the fiber-based HOM interferometer ceramic sleeve connectors between two fiber facets were used allowing to align fast and slow axis at the desired angle.	 
		}
	\end{figure}
	
	\newpage
	\section{S3: Optical set-up} 
	For all experiments, the sample is kept in a low-vibrations closed-cycle cryostat (attoDry800) at temperatures of $\sim$4.5~K. The cryostat is equipped with two optical windows allowing for access from both side and top of the sample. A spectroscopic setup consisting of two independent perpendicularly aligned optical paths is employed as shown schematically in Figure~\ref{fig:setup}. Additionally, the excitation path allows for the separate routing of two laser beams for the simultaneous excitation of two spots on the sample. For HOM and HBT experiments, a tunable Ti:Si picosecond pulsed laser is used. QDs embedded into the two input arms of the DC are excited from the top through a first microscope objective with x10 magnification and NA~=~0.26, while the emission signal from both DC output arms is detected simultaneously from the side facet of the sample with a second objective with x20 magnification and NA~=~0.4. Photoluminescence signal from both arms is then passed through a spatial filter (lenses L$_1$ and L$_2$) and polarization optics. For light polarization analysis, a half-wave plate (HWP) combined with a linear polarizer (LP) is used. The sample image plane is rotated from a horizontal into a vertical direction using a Dove prism, which allows simultaneous coupling signals from both DC output arms into the monochromator. The collected light is analyzed by a high-resolution monochromator equipped with a liquid nitrogen-cooled low-noise charge-coupled device detector (CCD), featuring a spectral resolution of $\sim$40~$\mu$eV. Taking advantage of the spatial separation of DC output arms, the spectrum from both WG arms is resolved spatially on the CCD camera. For HBT and HOM experiments, the monochromator is used as a spectral filter with 70~$\mu$eV width. Next, the signal from both arms is separated spatially using a knife-edge mirror and coupled into single-mode fibres interconnected with superconducting single-photon detectors (SSPD). The time-correlated measurements are acquired using a stand-alone time-tagger.
	
	\section{S4: Power-resolved PL} 
\begin{figure}[h]
		\includegraphics[width=6in]{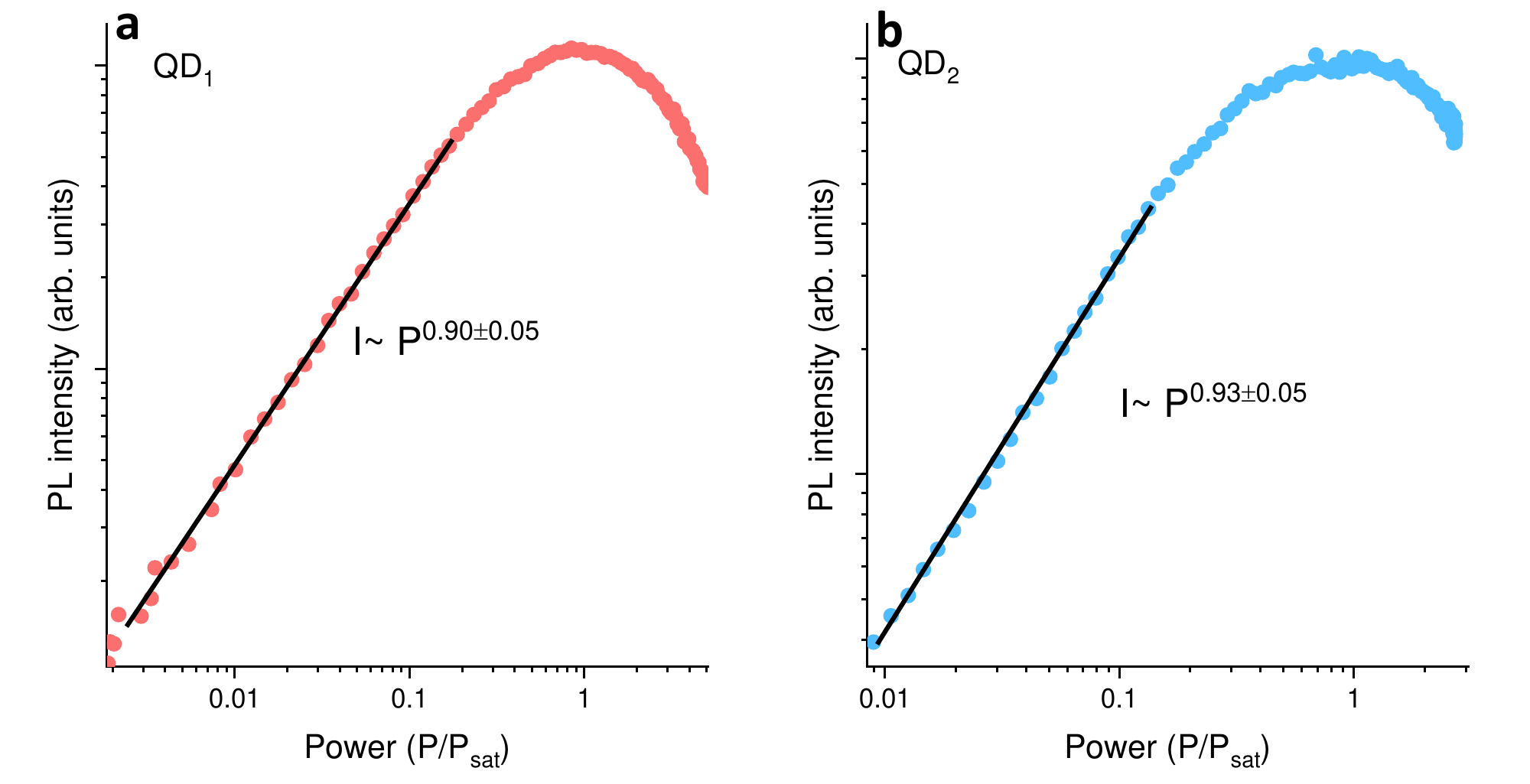}
		\caption{\label{fig:pow} Photoluminescence intensity vs incident excitation power. Solid red/blue curve: fit with a power function revealing linear dependence of the emission intensity on excitation power. 
		}
	\end{figure}

In Fig.~\ref{fig:pow} QD$_1$ and QD$_2$ emission intensity as a function of excitation power is shown. An almost linear dependence of the emission intensity on excitation power suggests that the analyzed lines originate from the recombination of neutral or charged excitonic complexes.
	
	\newpage
	\section{S5: Waveguide transmission losses} 
\begin{figure}[h]
		\includegraphics[width=6.0in]{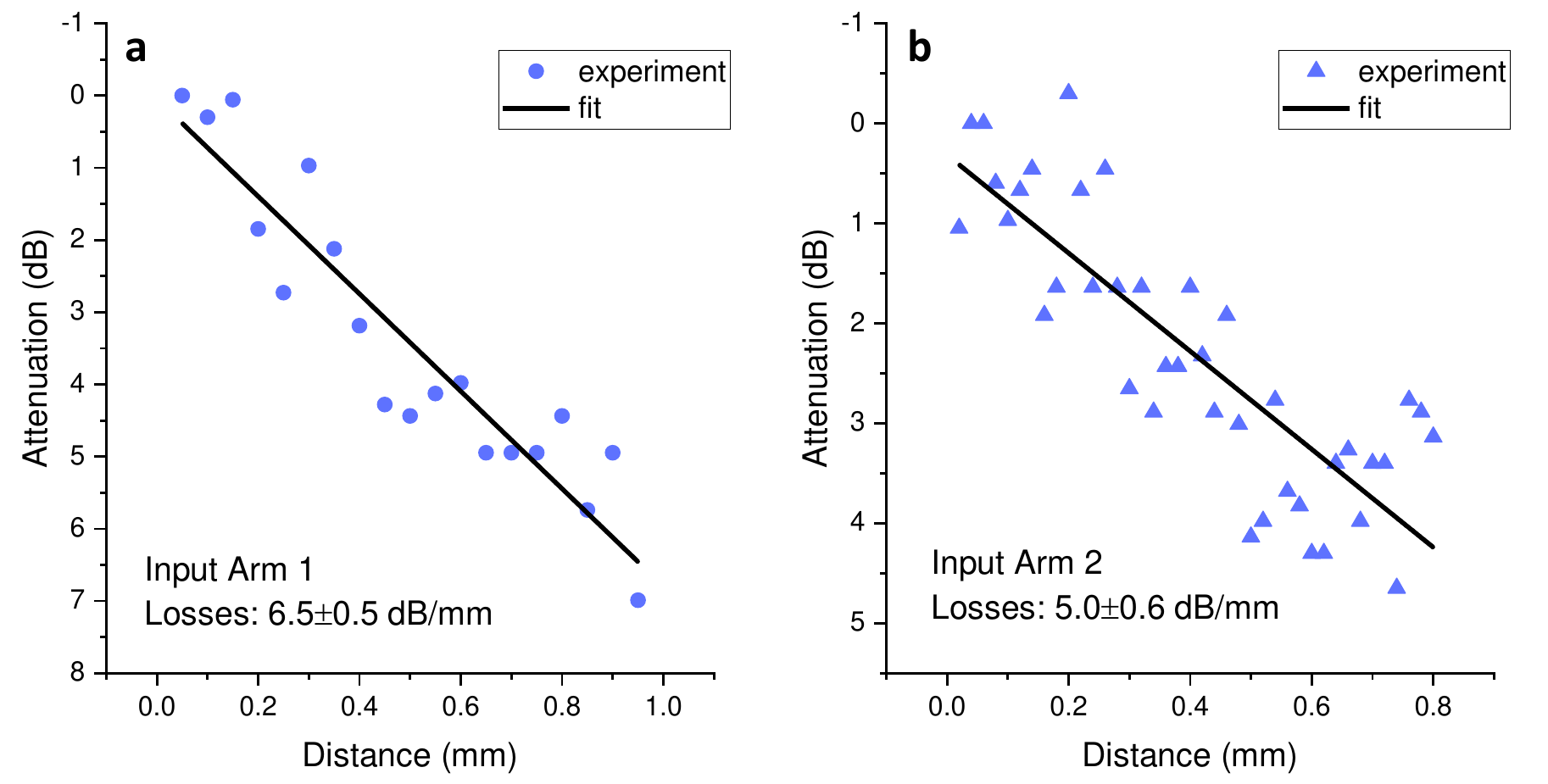}
		\caption{\label{fig:loss}Waveguides transmission losses. Attenuation of the side detected ensemble PL signal in function of the distance from DC bend regions. 
		}
	\end{figure}

To estimate the quality of the etched ridge waveguides, the optical WG transmission losses were determined. For that purpose, the sample was excited with very high pumping power, allowing us to observe spectrally broad QD ensemble emission. The beam spot was scanned along the DC input arm 1/2, and emission was detected from the side through the waveguide arm 1. Figure~\ref{fig:loss}a and b show the corresponding attenuation of the measured intensities at 890~nm plotted as a function of the distance to the DC bends for input arms 1 and 2, respectively. Input arm 1 exhibits transmission losses on the level of 6.5$\pm$0.5~dB/mm and arm 2 of 5.0$\pm$0.6 dB/mm. Waveguide transmission characteristics are limited by ridge sidewall imperfections, which could be potentially further improved by optimizing the etching process. 
	
	\newpage
	\section{S6: Polarization-resolved PL} 
\begin{figure}[h]
		\includegraphics[width=5in]{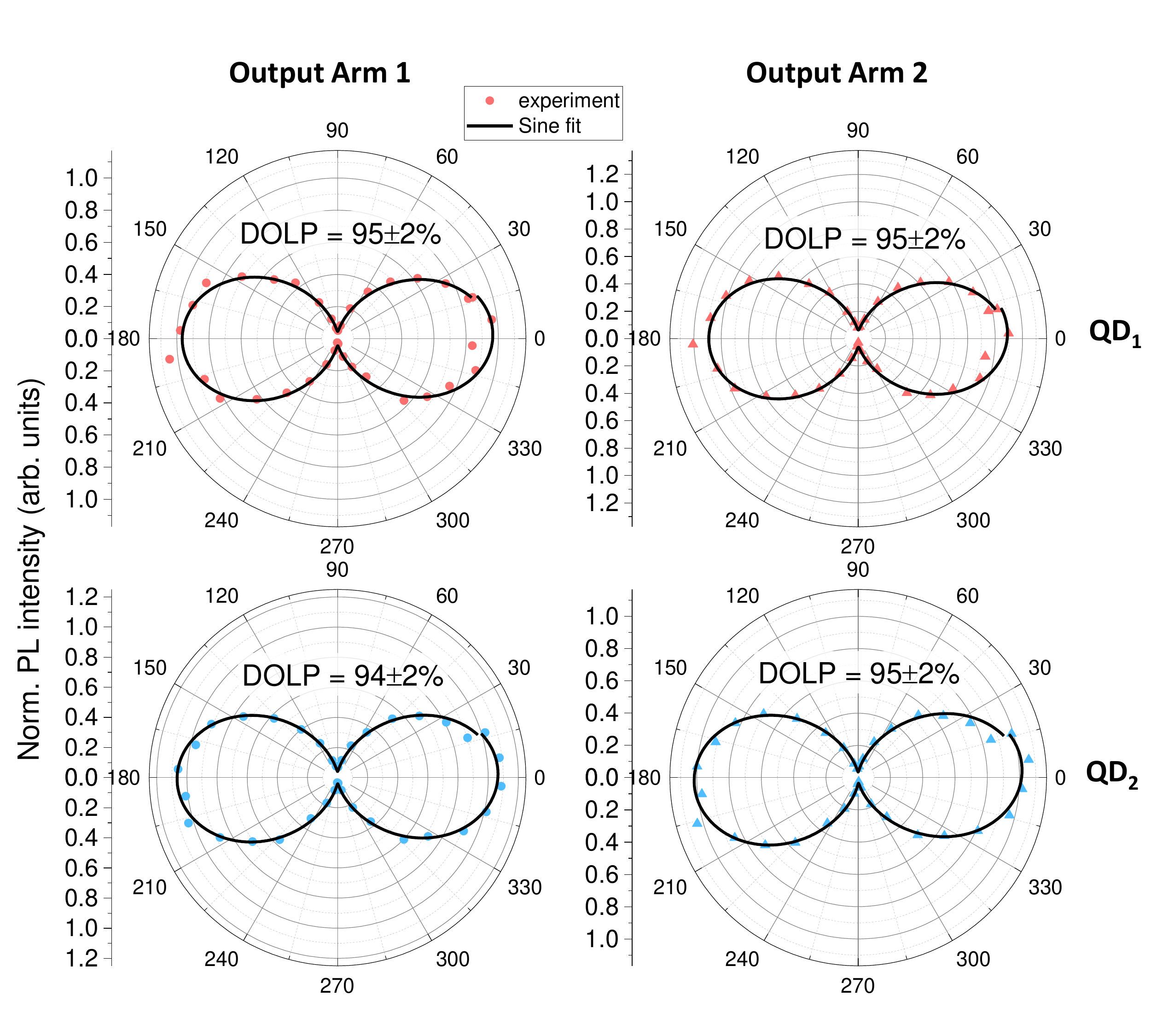}
		\caption{\label{fig:pol}Polarization characteristics of the QD$_1$ and QD$_2$ emission coupled to input arms of the DC and detected from side facet output arms 1 and 2. Both QDs PL emission is strongly linearly polarized with around 95\% degree of linear polarization.
		}
	\end{figure}

Side detected emission from both studied emission lines show a high degree of linear polarization (DOLP) of around 95\%, oriented in the sample plane as shown in Fig.~\ref{fig:pol}. A high DOLP and its direction are related to the QDs dipole moments, which are mainly in-plane oriented and thus emitted photons mostly couple to and propagate in the TE waveguide mode. It needs to be noted that high DOLP is maintained after passing the whole circuits consisting of bends, DC, and out-coupler, which could potentially spoil the detected polarization contrast. The same polarization level is observed for both output arms of the DC. 
	
	\section{S7: Emitters transition linewidths} 
\begin{figure}[h]
		\includegraphics[width=6.5in]{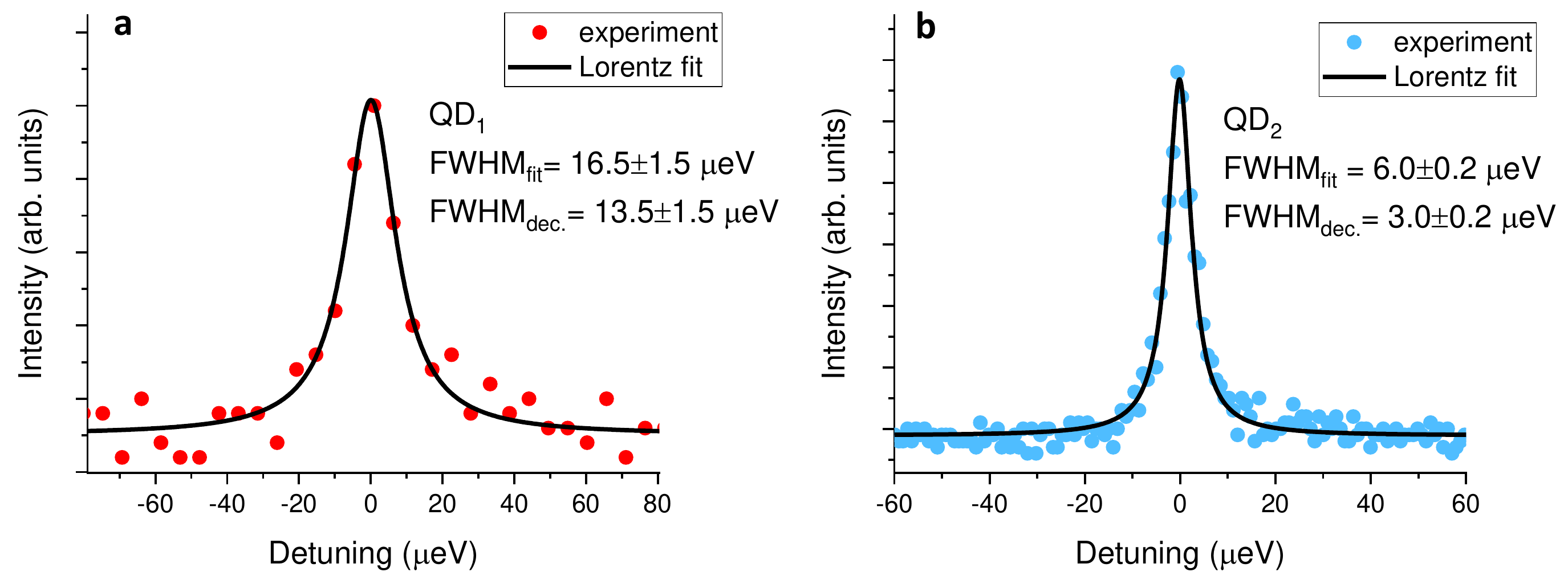}
		\caption{\label{fig:FP}A high-resolution PL spectrum of QD$_1$ and QD$_2$, obtained using a home-built Fabry-Perot scanning cavity with a resolution of 3~$\mu$eV (FWHM), and free spectral range of 140~$\mu$eV at 890~nm. Solid lines are fits with the Lorentz function. 
		}
	\end{figure}

Figure~\ref{fig:FP} shows a high-resolution spectrum of the QD$_1$ and QD$_2$ emission recorded under cw 660~nm excitation. Measurements are performed using a scanning Fabry-Perot interferometer with a 3~$\mu$eV Lorentzian profile linewidth. Both spectra are fit using Lorentzian functions with full-width at half-maximum (FWHM) of 16.5$\pm$2.5~$\mu$eV and 6.0$\pm$0.2~$\mu$eV, for QD$_1$ and QD$_2$, respectively (error related to fit precision). It can be shown that the convolution of two Lorentzian profiles with FWHM$_1$ and FWHM$_2$ is also a Lorentzian profile with broadening of FWHM$_1$+FWHM$_2$. Following the above, we can correct the recorded optical linewidths for the finite resolution of our setup by simply subtracting its linewidth. The deconvoluted linewidths are 13.5$\pm$2.5~$\mu$eV and 3.0$\pm$0.2~$\mu$eV for QD$_1$ and QD$_2$, respectively.
	
	\section{S8: Directional coupler characteristics} To extract the DC splitting ratio $r:t$ accounting for the different performance of both output arms due to the fabrication imperfections, the following procedure has been used. First, QD located in input arm 1 was excited, and PL signal from output arm 1 $I^{I1}_{O1}$ and arm 2 $I^{I1}_{O2}$ collected. Next, the same measurement has been repeated on QD located in arm 2, revealing $I^{I2}_{O1}$ and $I^{I2}_{O2}$. If the uneven outcoupling efficiency between the two output arms is quantified as a constant $x$ (transmission ratio between two arms), the following set of equations applies   
	\begin{equation}
	\begin{split}
	r+t=1, \\
	x\frac{r}{t} = \frac{I^{I1}_{O1}}{I^{I1}_{O2}}, \\ 
	x\frac{t}{r} = \frac{I^{I2}_{O1}}{I^{I2}_{O2}}. \\
	\end{split}
	\end{equation}
	Based on that, the DC splitting ratio $r:t$ can be derived, accounting for the imbalanced outcoupling
	\begin{equation}
	r:t = \sqrt{\frac{I^{I1}_{O1}}{I^{I1}_{O2}} \frac{I^{I2}_{O2}}{I^{I2}_{O1}}}.
	\end{equation}
	In the case of the investigated DC with $I^{I1}_{O1}/I^{I1}_{O2}$ of 51:49 and  $I^{I2}_{O2}/I^{I2}_{O1}$ of 46:54 we ended up with the r:t ratio of 48:52, which is very close to the desired 50:50.  
	
	\begin{figure}[h]
		\includegraphics[width=4.5in]{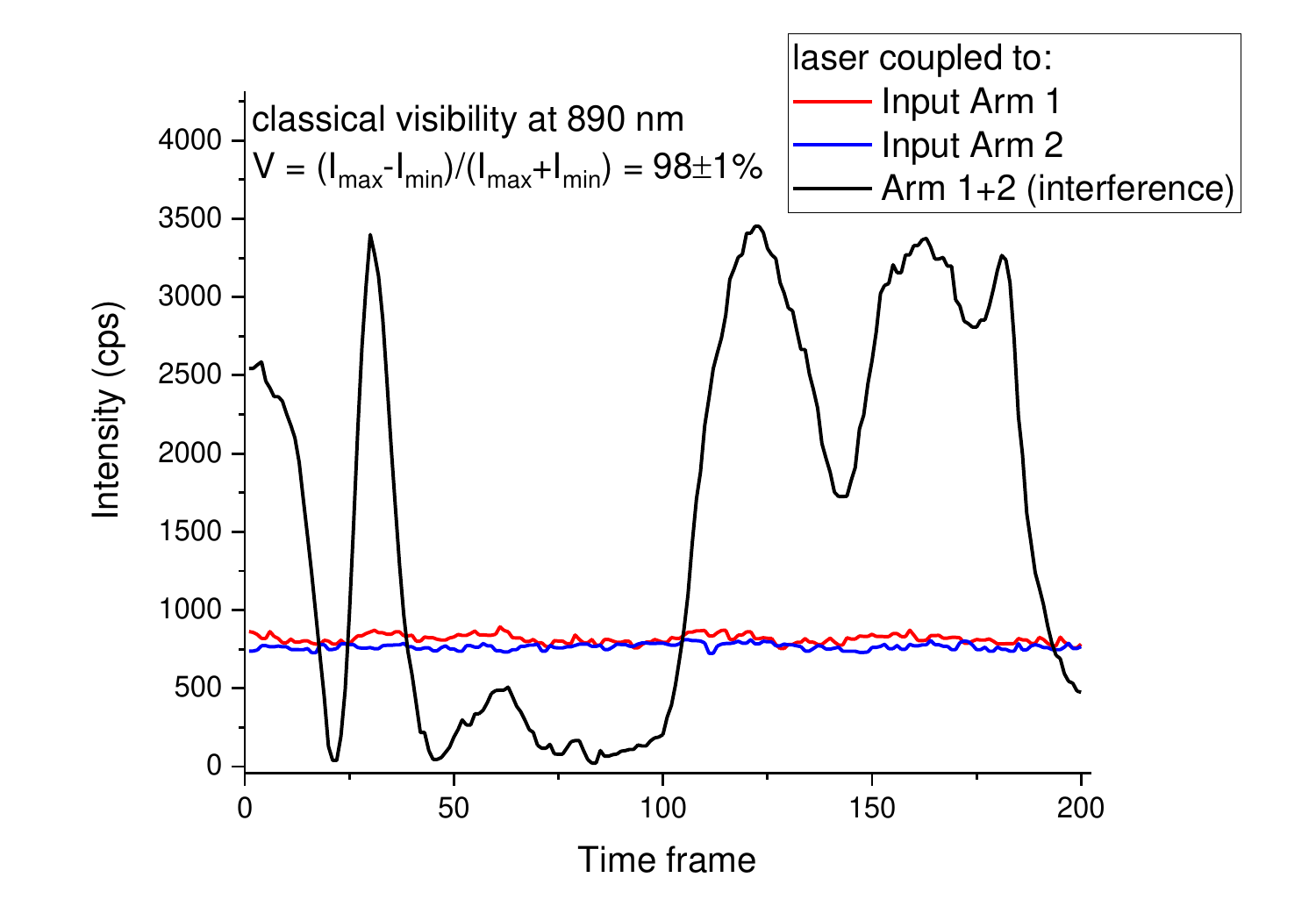}
		\caption{\label{fig:DC}Intensity fluctuations of the cw laser light transmitted simultaneously through two input arms of the DC. Fluctuations in time are related to the off-chip path-length difference fluctuations of signal interfered on the directional coupler. 
		}
	\end{figure}
	
	To test the classical visibility of the DC device, we simultaneously send cw laser light tuned to the energy of QDs transitions (890~nm), using circular reflectors placed on the ends of the input arms waveguides. The power of the laser coupled into both arms was adjusted, so the intensity from both input arms was the same. Next, we focused on the signal passing through the DC and out-coupled from one of the output arms. We observed intensity modulation in the function of time, related to the small path-length difference fluctuation, allowing us to record interference pattern as shown in Fig.~\ref{fig:FP}. Classical DC visibility was obtained by calculating the intensity contrast
	\begin{equation}
	V_{DC}=\frac{I_{max}-I_{min}}{I_{max}+I_{min}}.
	\end{equation} 
	In the case of the investigated DC device, a classical visibility of 98$\pm$1\% was extracted.

	\newpage
	\section{S9: Raw HOM correlation data} 
\begin{figure}[h]
		\includegraphics[width=6in]{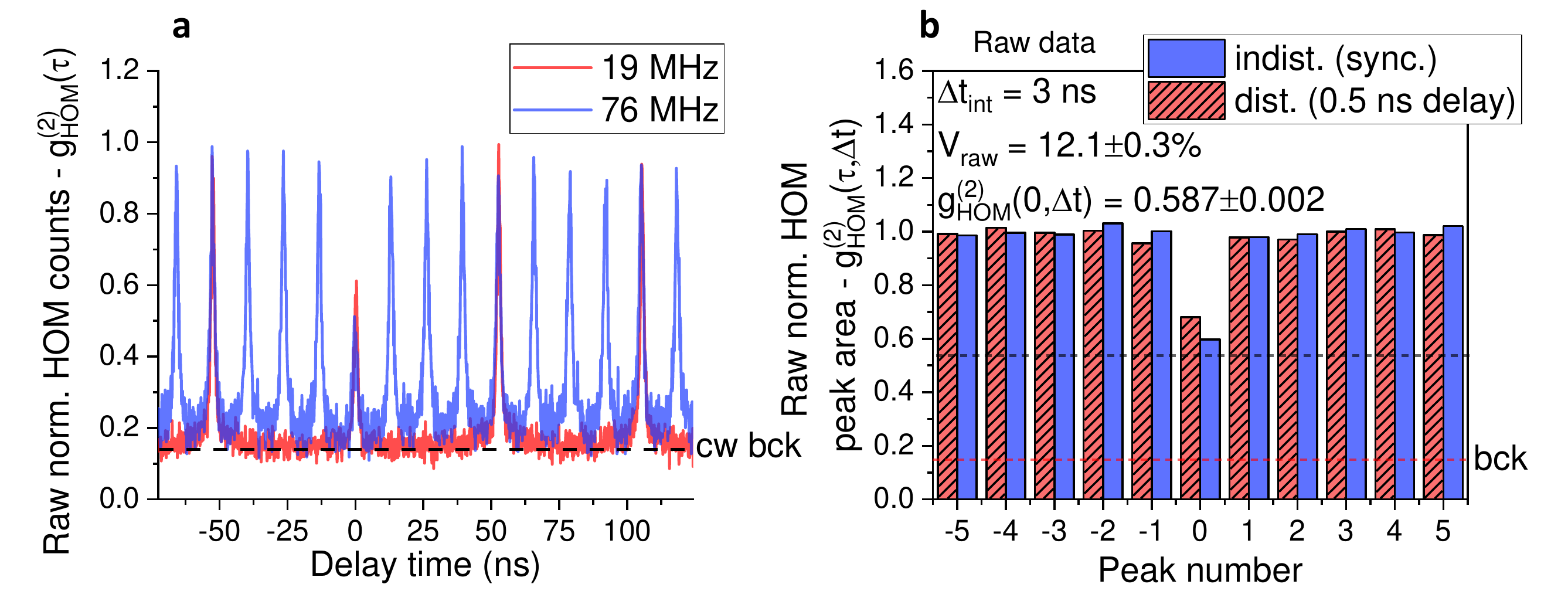}
		\caption{\label{fig:bck}a,~Two-photon Hong-Ou-Mandel interference measurement between QD$_1$ and QD$_2$ performed using on-chip beamsplitter showing the normalized coincidences versus the delay time with no background counts correction. An experiment performed under 76MHz and 19MHz pulsed laser repetition rate yields the same cw background. b,~Non-corrected integrated counts of the central eleven peaks (3~ns integration window) of the HOM correlation in case of synchronized (blue bars) and 0.5~ns delayed (red bars) photons from QD$_1$ and QD$_2$ under pulsed 76MHz excitation.
		}
	\end{figure}

Figure ~\ref{fig:bck}a shows a non-corrected two-photon Hong-Ou-Mandel interference experiment result between QD$_1$ and QD$_2$ performed under 76MHz and 19MHz pulsed laser repetition rate. In the case of both graphs, the same time-independent background offset is visible, corresponding to around 15\% of the normalized peak intensity. Since the background level is the same for the HOM graphs at different laser repetition rates, we exclude the emitters long-decay contribution to the observable background. We attribute the observed cw offset to the SSPDs dark counts (100-500~cps). Figure ~\ref{fig:bck}b shows the non-corrected normalized histogram of the central eleven peaks areas ($\Delta=$~3~ns integration window) of the HOM second-order cross-correlation in case of synchronized - indistinguishable (red bars) and 0.5~ns delayed - distinguishable (grey bars) photons. The non-corrected central peak area in case of synchronized photons is equal to $g^{(2)}_{HOM}(0,\Delta t)=$~0.587$\pm$0.002, while in case of unsynchronized photons $g^{(2)}_{HOM_{d}}(0,\Delta t)=$~0.680$\pm$0.002. In this case, the non-corrected two-photon interference visibility yields $V_{raw}=$~12.1$\pm$0.3\% in correspondence to 17.8$\pm$0.7\% visibility obtained for background-corrected graphs (see Fig.3c in the main text).
	
	\section{S10: Single photon emission under cw excitation } 
\begin{figure}[h]
		\includegraphics[width=6in]{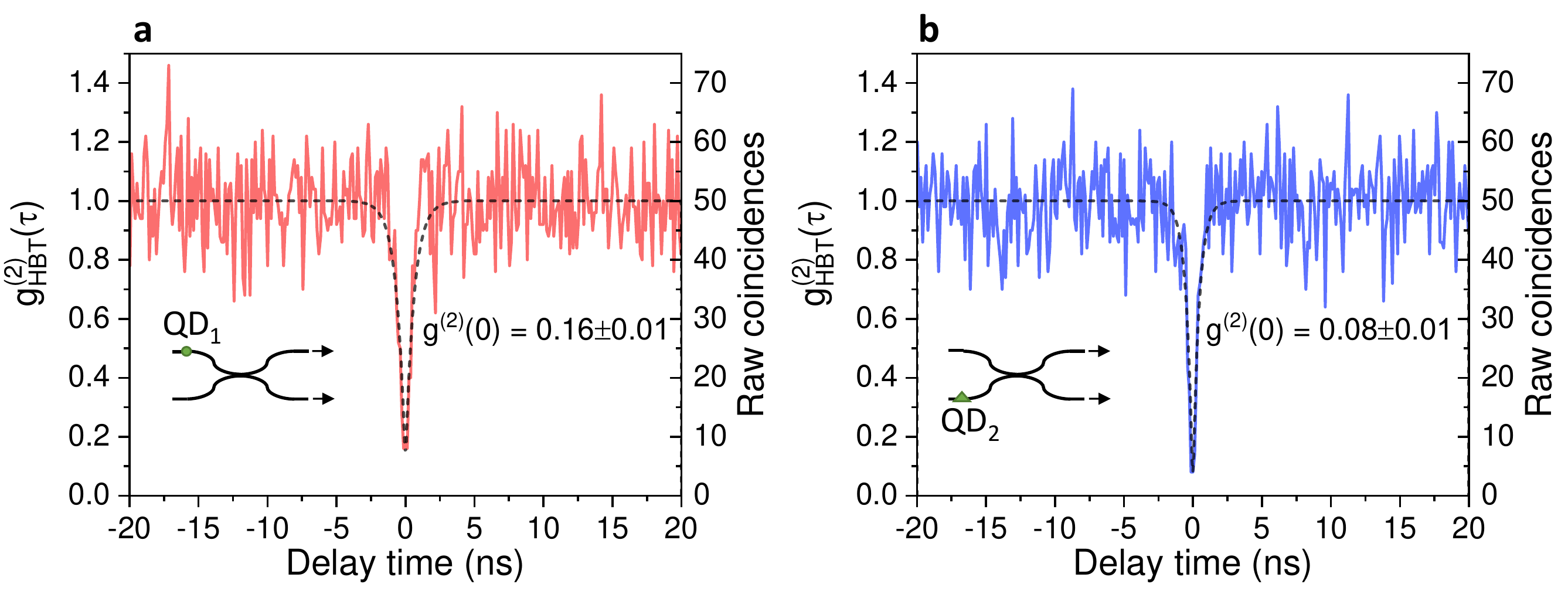}
		\caption{\label{fig:g2_cw}Second order auto-correlation histograms of a QD$_1$ and b QD$_2$ emission under non-resonant (660 nm) cw excitation. Data have been recorded in HBT configuration using an on-chip beamsplitter. Presented data are shown as measured with no background subtraction or other corrections.
		}
	\end{figure}

	In Figure~\ref{fig:g2_cw}, HBT second-order correlation histograms recorded under cw (660~nm) excitation for QD$_1$ and QD$_2$ are shown. The data in Fig.~\ref{fig:g2_cw} are fit with the function $g^{(2)}_{HBT}(\tau)=(1-g^{(2)}_{HBT}(0))exp(-|\tau|/\tau_d)$ convoluted with 80~ps width Gaussian instrumental response function, where $\tau$ is the delay time between detection events, $g^{(2)}_{HBT}(0)$ is the probability of two-photon emission events, $\tau_d$ is decay time constant corresponding to the sum of the spontaneous emission rate 1/$T_1$ and pump rate $G$ of the source.  
	
